\documentclass[pre,twocolumn,superscriptaddress,nofootinbib,amsmath,amssymb,longbibliography]{revtex4-1}
\usepackage{graphicx}
\usepackage{float}
\usepackage[usenames,dvipsnames]{xcolor}
\usepackage[colorlinks=true,citecolor=Blue,linkcolor=RubineRed,urlcolor=Blue]{hyperref}
\begin{document}

\title{
Transition from Susceptible-Infected to Susceptible-Infected-Recovered Dynamics 
in a Susceptible-Cleric-Zombie-Recovered Active Matter Model 
}
\author{A. Lib{\' a}l}
\affiliation{Mathematics and Computer Science Department, Babe{\c s}-Bolyai University, Cluj-Napoca 400084, Romania}
\author{ P. Forg{\' a}cs}
\affiliation{Mathematics and Computer Science Department, Babe{\c s}-Bolyai University, Cluj-Napoca 400084, Romania}
\author{ {\' A}. N{\' e}da}
\affiliation{Mathematics and Computer Science Department, Babe{\c s}-Bolyai University, Cluj-Napoca 400084, Romania}
\author{C. Reichhardt}
\affiliation{Theoretical Division and Center for Nonlinear Studies,
Los Alamos National Laboratory, Los Alamos, New Mexico 87545, USA}
\author{N. Hengartner}
\affiliation{Theoretical Division and Center for Nonlinear Studies,
Los Alamos National Laboratory, Los Alamos, New Mexico 87545, USA}
\author{C. J. O. Reichhardt$^*$}
\affiliation{Theoretical Division and Center for Nonlinear Studies,
Los Alamos National Laboratory, Los Alamos, New Mexico 87545, USA}

\date{\today}

\begin{abstract}
The Susceptible-Infected (SI) and Susceptible-Infected-Recovered (SIR) models provide two distinct representations of epidemic evolution, distinguished by the lack of spontaneous recovery in the SI model. Here we introduce a new active matter epidemic model, the ``Susceptible-Cleric-Zombie-Recovered'' (SCZR) model, in which spontaneous recovery is absent but zombies can recover with probability $\gamma$ via interaction with a cleric. Upon interacting with a zombie, both susceptibles and clerics can enter the zombie state with probability $\beta$ and $\alpha$, respectively. By changing the intial fraction of clerics or their healing ability rate $\gamma$, we can tune the SCZR model between SI dynamics, in which no susceptibles or clerics remain at long times, and SIR dynamics, in which no zombies remain at long times. The model is relevant to certain real world diseases such as HIV where spontaneous recovery is impossible but where medical interventions by a limited number of caregivers can reduce or eliminate the spread of infection.
\end{abstract}
\maketitle

\section{Introduction}
\begin{figure}
\includegraphics[width=0.5\textwidth]{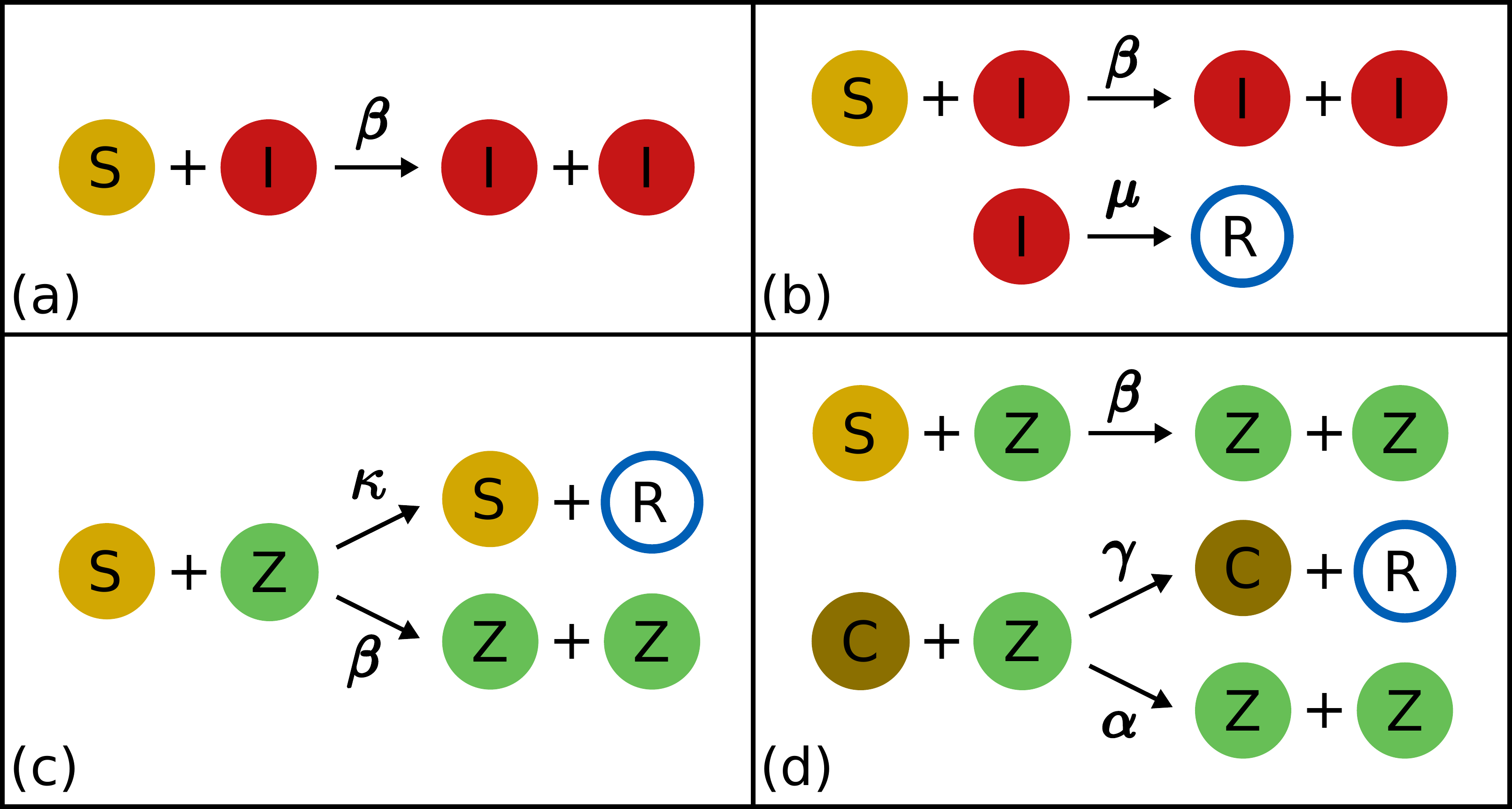}
\caption{
(a) In the SI model, there is no spontaneous recovery, and
susceptibles ($S$, yellow) that come into contact with
infectives ($I$, red) become infected with probability $\beta$. 
(b) The SIR model adds
a spontaneous recovery process in which
an $I$ transitions to recovered ($R$, blue open circle) at a rate $\mu$.
(c) In the Living-Zombie-Recovered model introduced in Ref.~\cite{Alemi15},
a zombie ($Z$, green) interacting with $S$
recovers with probability $\kappa$ and turns the $S$ into $Z$ with
probability $\beta$.
(d) In our SCZR model, we divide the susceptible population into $S$ and clerics
($C$, brown). $Z$ can only recover
when in contact with $C$ with probability $\gamma$, but interaction with
$Z$ causes $S$ to turn into $Z$ with probability $\beta$ and $C$ to turn into
$Z$
with probability $\alpha$.
}
\label{fig:1}
\end{figure}

Understanding the propagation of infectious diseases is an intensely
studied issue, and a variety of different 
epidemic models and methods to simulate the spread of disease have
been developed
\cite{Kermack27,Bailey75,Hethcote00,Martcheva15}.  Two of the
most widely used
disease propagation models
are the Susceptible-Infected (SI) and Susceptible-Infected-Recovered
(SIR) models
\cite{Kermack27,Bailey75,Hethcote00,Martcheva15}.
In the SI model, illustrated in Fig.~\ref{fig:1}(a),
there are only susceptibles ($S$)
and infectives ($I$) present. There is no spontaneous recovery,
and the model contains only a single probability $\beta$
for an $S$ to transform to an $I$.
As shown in Fig.~\ref{fig:1}(b),
the SIR model adds
a spontaneous recovery process with rate $\mu$ for an $I$
to become recovered ($R$).
A key difference between the SI and SIR models is that
in the SI model the amount of $S$ present drops to zero at long times,
but in the SIR model
the amount of $I$ present drops to zero.
A wide range of diseases can be described using these two models.
Diseases with lifelong transmittivity
and no recovery are captured by the SI model,
while situations where reinfection is impossible but spontaneous recovery occurs
can be represented with the SIR model.
Numerous variations of the SI and SIR models have been considered 
over the years \cite{Bailey75,Hethcote00,Martcheva15,Bjornstad20},
including epidemic spreading on networks \cite{PastorSatorras15}, memory 
effects \cite{Bestehorn22}, adding vaccination \cite{Gao07},
spatial heterogeneity \cite{Keeling99,Tildesley09}, social distancing
\cite{teVrugt20}, diffusion \cite{Polovnikov22},
and models that include
details on mobility patterns in attempts to more accurately portray
real world epidemics
\cite{Eubank04,Germann06}.

Despite the large number of models that have been explored,
we did not find any descriptions
of a model 
in which a transition from SI to SIR behavior naturally emerges.
Such transitions could arise for certain types of 
infectious disease where spontaneous recovery does not occur but
where direct medical intervention can result in recovery
or a reduced rate of infectiousness.
For example, in the human immunodeficiency virus (HIV),
an untreated patient
remains contagious,
but when appropriate medical interventions are applied,
the patient becomes effectively cured and has a
rate of infectiousness that drops dramatically or even reaches zero.
In such cases, if there is an insufficient supply of resources or
treating agents (doctors), the course of the epidemic will follow
the SI model, but if there are ample resources or treating agents,
the epidemic will instead fall in the SIR regime.

Standard SI and SIR type models assume homogeneous mixing of
infectious and susceptible individuals, either across the entire population
or within strata.  For many diseases, that assumption is known to fail
and in Refs.~\cite{Burr2008,Grossmann2021.03.25.21254292},
the impact 
of the failure of the homogeneity assumption is studied.
In our previous work \cite{Forgacs22}, we showed 
that a run-and-tumble active matter model combined with SIR
dynamics produces different regimes of behavior when quenched
disorder is introduced, due to the lack of homogeneous mixing
in the system.
For low infection rates, the quenched disorder strongly affects the
duration of the epidemic as well as the final epidemic size or 
fraction of $S$ that
survive to the end of the epidemic.
When the infection rate is high, the quenched disorder has little
impact and the epidemic propagates as waves through the system.

The term ``active matter'' encompasses self driven systems such as
an assembly of self-motile particles that undergo contact interactions
with each other \cite{Marchetti13,Bechinger16}.
In our previous work \cite{Forgacs22}, we considered run-and-tumble
particles moving in two dimensions and subjected to rules of how an
infection spreads when a contact interaction occurs between an
$S$ and an $I$ particle.
Active matter systems are attractive for epidemic modeling
since they allow real world effects such as spatial heterogeneity to
be incorporated easily because density heterogeneities arise naturally
from the interactions among the particles,
and there have now been several studies
in which active matter is used to study epidemics 
\cite{Paoluzzi20,Norambuena20,Zhao22}.
There have also
been several experimental realizations of active matter 
systems that can mimic social dynamics through the
activity and tracking of individual active particles, 
so the type of active matter epidemic systems
we consider here should be feasible to create experimentally 
\cite{Lavergne19,Bauerle20}.

Here we introduce a new model for epidemic spreading
featuring
multiple susceptible species and no spontaneous
recovery, and show that in this model, an easily tunable transition between
SI and SIR behavior occurs.
We specifically consider a modification of the
Susceptible-Zombie-Removed (SZR) model
proposed by Alemi {\it et al.} \cite{Alemi15}.
Figure~\ref{fig:1}(c) shows the dynamics of the SZR
model. Unlike the SIR model, the SZR model has no spontaneous recovery.
Instead, when an $S$ and a zombie ($Z$) interact, the $Z$ transitions to
recovered ($R$) with probability $\kappa$, while the $S$ transitions
to $Z$ with probability $\beta$.
In our
modification of the model, there is again no spontaneous recovery, but we break
the susceptible population into two portions: susceptibles ($S$) and
clerics ($C$). As illustrated in Fig.~\ref{fig:1}(d),
when an $S$ interacts with a $Z$, the $S$ becomes a $Z$
with probability $\beta$,
as in the SZR model; however, the $S$ cannot cause the $Z$ to recover. Instead,
only an interaction between a $C$ and a $Z$ can cause the $Z$ to recover with
probability $\gamma$, while with probability $\alpha$, the $C$ becomes a $Z$.
We call this the
Susceptible-Cleric-Zombie-Removed or
``SCZR'' model. Although, as in Ref.~\cite{Alemi15}, we
have placed the model in a zombie framework, the model can be
rephrased in terms of
certain real world diseases such as HIV which, 
if left untreated, confer a lifelong ability to infect;
however, under medical treatment from a health care provider,
the infection rate can be reduced or dropped to zero,
resulting in an effectively recovered individual.
In this case, the zombie class
would be simply be labeled as infected ($I$) while the cleric class
would represent some form of health care provider or medical resources.
As we show below, the SCZR model exhibits SI behavior when the initial
fraction of $C$ or the healing rate $\gamma$ is low, since in this case
the $Z$ wipe out both the $C$ and the $S$ so that a finite fraction
of $Z$ remain at the end of the epidemic.
In contrast, when the initial fraction of $C$ or the healing rate $\gamma$ is
high enough, the $C$ are able to eliminate the $Z$ so that a finite fraction of
$S$ and $C$ remain at the end of the epidemic, which is behavior associated with
an SIR model.

\section{Modeling and characterization of the SCZR dynamics}

We consider a two-dimensional assembly
of $N=5000$ run-and-tumble active particles 
in a system of size $L \times L$ where $L=200.0$ and
where there are periodic boundary
conditions in both the $x$ and $y$ directions. The motion of the
particles is obtained by integrating
the following overdamped equation of motion in discrete time:
\begin{equation} 
\alpha_d {\bf v}_{i}  =
{\bf F}^{dd}_{i} + {\bf F}^{m}_{i} \ .
\end{equation}
Here ${\bf v}_{i} = {d {\bf r}_{i}}/{dt}$ is the velocity and  
${\bf r}_{i}$ is the position of particle $i$,
and the damping constant $\alpha_d = 1.0$.
The interaction between two particles,
each of radius $r_a=1.0$, is modeled with a harmonic
repulsive potential
${\bf F}^{dd}_{i} = \sum_{i\neq j}^{N}k(2r_{a} - |{\bf r}_{ij}|)\Theta( |{\bf r}_{ij}| - 2r_{a}) {\hat {\bf r}_{ij}}$, where $\Theta$ is the Heaviside step function, ${\bf r}_{ij} = {\bf r}_{i} - {\bf r}_{j}$, $\hat {\bf r}_{ij}  = {\bf r}_{ij}/|{\bf r}_{ij}|$, and the repulsive spring force constant is
$k = 20.0$.

Each particle is subjected to an active motor 
force ${\bf F}_i^m=F_{M}{\bf \hat{m}}_i$ of magnitude $F_M$ applied in a randomly chosen direction ${\bf \hat{m}}_i$
during a continuous run time of $\tau_{l}\in [1.5\times 10^4$, $3.0 \times 10^4]$ before 
instantaneously changing to a new randomly chosen direction.
This type of run-and-tumble dynamics of active particles has
been used extensively
to model active matter systems \cite{Marchetti13,Bechinger16,Cates15},
active ratchets \cite{Reichhardt17a}, active jamming \cite{Reichhardt14} and
motility induced phase separation \cite{Cates15,Sandor17a}. 
In another version of active matter, the particles 
undergo driven diffusion; however, many of the generic phases
are the same for both run-and-tumble and
driven diffusive active matter \cite{Cates15,Cates13},
so we expect that our results 
will also be relevant to driven diffusive systems.
For sufficiently large density or activity,
both run-and-tumble and driven diffusive active particles
begin to exhibit self-clustering,
leading to what is known as motility-induced phase separation (MIPS)
\cite{Marchetti13,Bechinger16,Cates15,Fily12,Redner13,Palacci13,Buttinoni13}. 

We select the run length range
and motor force value such that the system
is in the MIPS regime, and thus creates large connected active clusters
similar to those employed in our previous active matter epidemic 
model \cite{Forgacs22}.
Each particle tracks which one of the four possible states,
$S$, $Z$, $C$ or $R$, it is currently occupying.
These states are linked together by the
following equations:
\begin{eqnarray} 
	dS &&= -\beta SZ \label{eq:SCZR1}\\
	dZ &&= \alpha CZ + \beta SZ - \gamma CZ \label{eq:SCZR2}\\
	dC &&= -\alpha CZ \label{eq:SCZR3}\\
	dR &&= \gamma CZ \label{eq:SCZR4}\ .
\end{eqnarray}
According to these equations,
when an $S$ particle encounters a $Z$ particle, it changes its label to $Z$ 
with rate $\beta$.  More interestingly, when a C and Z particle come in
contact, a change in state occurs with rate $\alpha+\gamma$.
For interactions in which a state change occurs, with probability 
$\alpha/(\alpha+\gamma)$ the $C$ particle becomes a $Z$, and with probability 
$\gamma/(\alpha+\gamma)$, the $Z$ morphs into $R$.  Our simulation discretizes
time in $\Delta$-sized steps, and in the above dynamic, rates are changed into
probabilities.  Specifically, the probability that
an $S$ particle in contact with a $Z$ particle
morphs into a $Z$ particle is $1-e^{-\Delta \beta}$.  Similarly, 
the probability that
a change occurs during a $Z$ and $C$ particle encounter is $1-e^{-\Delta(\alpha-\gamma)}$.
The probability of transitions from $C$ to $Z$ and $Z$ to $R$
remains unchanged.

If at a given time step an $S$ particle is 
in contact with multiple $Z$ particles, or a
$Z$ particle is in contact with multiple $C$ or $S$ particles, 
every possible pair interaction is computed
independently using the unmodified states
of all particles, and the state of each particle is updated simultaneously
at the end of the computation when we apply all
$S\rightarrow Z$, $Z\rightarrow R$, and $C\rightarrow Z$ transitions.
There are no concurrency issues since
each type of particle can undergo only one type of transition.

The $R$ state is absorbing since the $R$ particles
experience no further state transitions,
but there is no mechanism to replenish the initial pool of
either $S$ or $C$
particles. The epidemic ends when either there are no more $S$
and $C$ particles or there are no more $Z$ particles.
Therefore, there are only two possible types of final
state for the SCZR model: an SI-like situation in which all $S$ and $C$
particles have been transformed into $Z$ and $R$ particles
(indicating that the zombies or the clinical cases prevail),
and an SIR-like situation in which all $Z$ particles have been extinguished
by becoming $R$ particles
(indicating that the medical community prevails and no
zombies or clinical cases remain).
While the time $t_d$ to reach the final state is finite,
we observe in simulations that $t_d$ can become very long
because,
in order for the epidemic to come to a conclusion,
it is necessary for the remaining $S$ and $C$ or the remaining $Z$ particles to
come into contact with $Z$ or $C$ particles, respectively.

We initialize the system by randomly placing the particles at non-overlapping
positions in the sample.
Initially all of the particles are set to the S state. We allow the system to
evolve for $5 \times 10^5$ simulation time steps until 
a large MIPS cluster
emerges, and we define this state to be the $t=0$ condition.
We then randomly select five particles and change their state to $Z$.
We choose five particles rather than one particle in order to lower the
probability of a failed outbreak.
We also randomly select
a fraction
ranging from $10\%$ to $100\%$ of the $S$ to
change into $C$.
The system continues to evolve under both the motion of the particles and the reactions between states $S$, $C$, $Z$, and $R$
until there are either no $S$ or $C$ particles or there are no $Z$ particles,
indicating that further epidemiological change is impossible.
We consider different values of $\alpha$, $\beta$, and $\gamma$
in addition to varying the fraction of $C$ in the initial population.

\section{Results}

\begin{figure}
\includegraphics[width=0.5\textwidth]{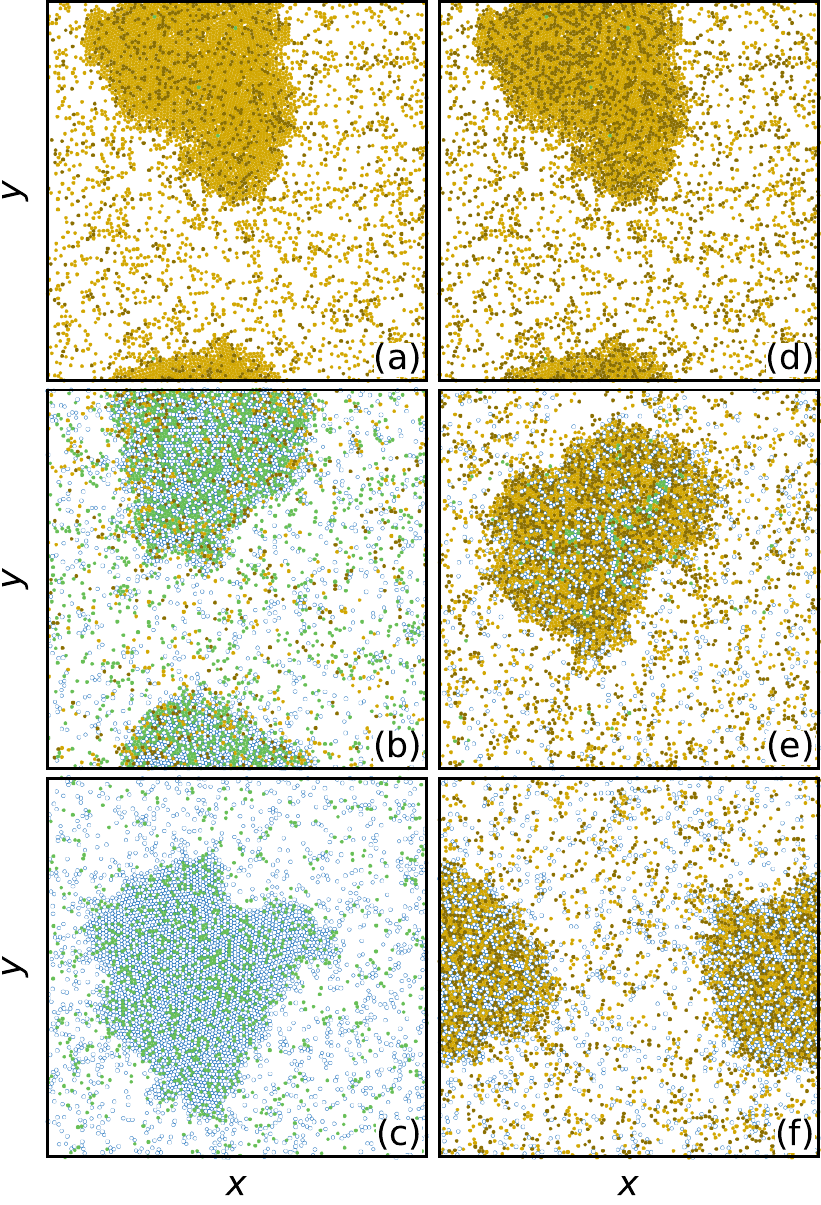}
\caption{
Snapshots of the time evolution of the SCZR system for
$\alpha=5\times 10^{-6}$, $\beta=1 \times 10^{-5}$, and
$\gamma=1.9\times 10^{-5}$.
Yellow disks are susceptibles ($S$), brown disks are clerics ($C$), green disks are zombies ($Z$), and open blue circles are recovered ($R$). (a,b,c) are for an
initial cleric fraction of $c_0=0.2$, 
and (d,e,f) are for
$c_0=0.4$.
(a,d) The $t=0$ moment where the MIPS cluster is present.
(b,e) The peak of the zombie outbreak, which occurs at
$t=4.85\times 10^5$ in (b) and at $t=9.67\times10^{5}$ in (e).
(c,f) The final state, which is reached at $t=1.606\times10^{6}$ in (c)
and $t=2.277\times10^{6}$ in (f). (a,b,c) show an SI-like evolution in which
all $S$ and $C$ are eliminated in the final state, while (d,e,f) show
an SIR-like evolution in which all $Z$ are eliminated in the final state.
}
\label{fig:2}
\end{figure}

In Figure \ref{fig:2} we illustrate the spatial 
evolution of our system under the SCZR model   
at fixed $\alpha=5\times10^{-6}$, $\beta=1 \times 10^{-5}$ and $\gamma=1.9\times10^{-5}$.
For Fig.~\ref{fig:2}(a,b,c), the initial fraction of $C$ is
$c_0\equiv C(t=0)/N=0.2$, and over
time we find an SI-like behavior in which
the zombie outbreak prevails and the populations of $S$ and $C$ drop to zero.
When $c_0$ 
is raised to $c_0=0.4$,
Fig.~\ref{fig:2}(d,e,f) shows an
SIR-like behavior in which recovery prevails and the
population of $Z$ drops to zero.
The initial condition of the MIPS cluster is identical for the two
cases, and the motion of the particles
is not influenced by their epidemiological state.
The peak of the zombie outbreak is shown in
Figs.~\ref{fig:2}(b) and \ref{fig:2}(e),
and the particle positions are different for the two cases only because
the peak in Fig.~\ref{fig:2}(e) occurs at a later time
of $t=9.67\times10^{5}$ compared
to the peak in Fig.~\ref{fig:2}(b),
which falls at $t=4.85\times 10^{5}$. In general we find that
the progression of an SIR-like epidemic
is significantly slower than that of an SI-like epidemic.
The end state of the epidemic is illustrated in Fig.~\ref{fig:2}(c)
when the last $C$ is eliminated after
a time of $t=1.606\times10^6$, and in Fig.~\ref{fig:2}(e)
when the last $Z$ is eliminated after a time
of $t=2.277\times10^6$.
%From the 3660 runs used to make the time histogram of the SI regime,
%in 2861 cases the S get eliminated first, but in the other 799 cases
%all the C get eliminated first.
In the well-mixed mean field limit, when
$\beta>\alpha$ we would expect that all of the $S$ are eliminated prior to the
elimination of the last $C$ for the $c_0=0.2$ system.
In practice, due to the heterogeneity of our system, we found
that out of all the SI simulations we considered,
the $S$ were eliminated prior to the $C$ 78\% of the time,
and the $C$ were eliminated prior to the $S$ 22\% of the time.

\begin{figure}
\includegraphics[width=0.5\textwidth]{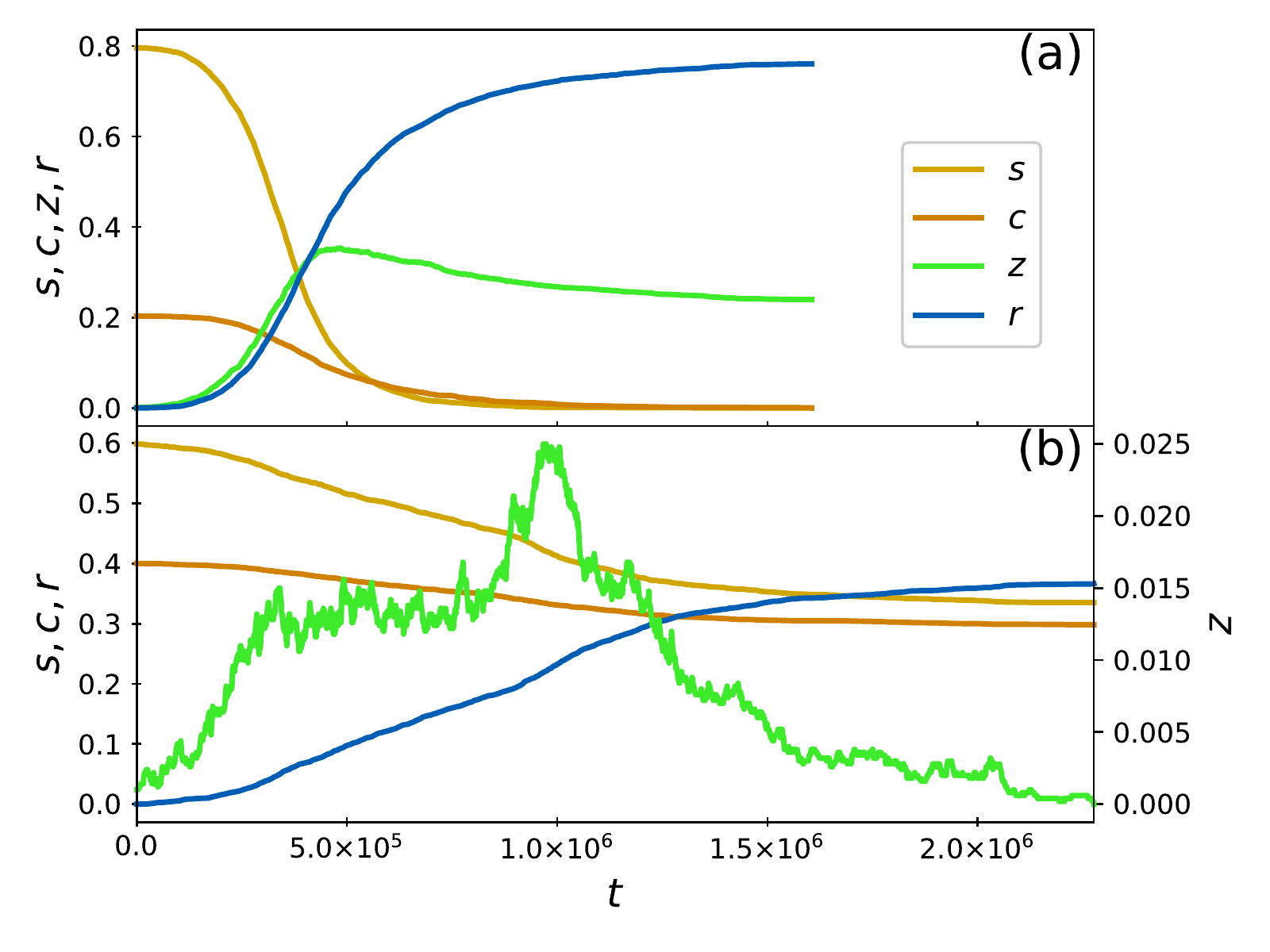}
\caption{
Epidemic curves for the individual runs
illustrated in Fig.~\ref{fig:2} with  
$\alpha=5\times10^{-6}$, $\beta=1\times 10^{-5}$ and $\gamma=1.9\times10^{-5}$
showing the fractions of susceptible $s$ (yellow), cleric $c$ (brown),
zombie $z$ (green), and recovered $r$ (blue) particles versus time
$t$.
(a) SI-like progression at
$c_0=0.2$ corresponding to the system 
in Fig.~\ref{fig:2}(a,b,c). Here, $s=c=0$ in the final state.
(b) SIR-like progression at
$c_0=0.4$ corresponding to the
system in Fig.~\ref{fig:2}(d,e,f).
The value of
$z$ is plotted on a separate $y$ axis for better visibility.
In the final state, $z=0$.
}
	\label{fig:3}
\end{figure}

In Fig.~\ref{fig:3}(a) we plot the epidemic
curves $s=S/N$, $c=C/N$, $z=Z/N$, and $r=R/N$ versus simulation time for
the $c_0=0.2$ system in the SI regime from Fig.~\ref{fig:2}(a,b,c).
At first, $r$ and $z$ increase at roughly the same rate until $z$ passes
through a local peak. Meanwhile,
since $\beta>\alpha$, $s$ decreases more rapidly than $c$,
and at longer times $z$ undergoes
a modest decrease from its peak value so that, at the end of the epidemic,
$s=0$, $c=0$, $z=0.25$, and $r=0.75$.
Figure~\ref{fig:3}(b) shows the epidemic curves for the
SIR regime with $c_0=0.4$ from Fig.~\ref{fig:2}(d,e,f).
Here the evolution
to the final state occurs much more slowly,
and in order to show the behavior of $z$ clearly we plot $z$ on a
separate $y$ axis scale, which is why the curve has a noisy appearance.
Both $s$ and $c$ decrease with time, but after passing through a peak,
$z$ drops to $z=0$ at the end of the epidemic while the values of $s$,
$c$, and $r$ all remain finite.
At late times during the epidemic in Fig.~\ref{fig:3}(b),
where all of the epidemic curves
become relatively flat, a strongly stochastic process occurs in which
the surviving $C$ and $Z$ need to come into contact with each other
in order to end the epidemic.
Since the motion of both $C$ and $Z$ is diffusive in nature, this slows
the progression of the epidemic and introduces more stochasticity.
For late times in Fig.~\ref{fig:3}(a), as the surviving $Z$
transform the remaining $C$ into $Z$, $z$ increases with each
transformation and so there is a higher probability of making contact
with the remaining $C$, shortening the epidemic. In contrast,
for late times in Fig.~\ref{fig:3}(b), the surviving $C$ transform
the remaining $Z$ into $R$, which are epidemiologically inert, so there
is no increase in $c$ with each transformation
and the total duration $t_d$ of the epidemic is
longer.

We next consider how changing the values 
of the model parameters $c_0$, 
$\alpha$, $\beta$, and $\gamma$ affects the epidemic outcomes. 
To characterize the outcome of a given
simulation, we introduce the quantity 
\begin{equation}
\upsilon= (s_f + c_f)/(s_0 + c_0),
\end{equation}
where $s_0=S(t=0)/N$ is the initial fraction of susceptibles,
$s_f=S(t=t_d)/N$ is the final fraction of susceptibles at time
$t=t_d$ equal to the duration of the epidemic, and
$c_f=C(t=t_d)/N$ is the final fraction of clerics.
Using $\upsilon$ we can determine
what fraction of the initial population of $S$ and $C$
survive the epidemic.
In the SI-like regime, $\upsilon=0$, and
in the SIR-like regime, $\upsilon$ remains finite.

From an epidemiological point of view, $\upsilon$ gives an indication
of how effective the medical intervention by the clerics is at
suppressing the epidemic. High values of $\upsilon$ are desirable
since this indicates that a smaller fraction of the population caught
the disease.
For any individual simulation with a given set of
parameters, it is possible to have either SI or SIR behavior emerge due
to the stochasticity, so we average $\upsilon$ over an ensemble of
50 runs for each parameter choice, where each run has a different
random seed for the initial particle positions and placement of $Z$ and
$C$ particles.
When $\langle \upsilon\rangle$ remains high,
the SIR behavior is dominant and the $Z$ are usually eliminated from the system,
while 
when $\langle \upsilon\rangle$ becomes small, the SI behavior is
dominant and the $S$ and $C$ are usually eliminated from the system so that
the zombies prevail.

\begin{figure}
\includegraphics[width=0.5\textwidth]{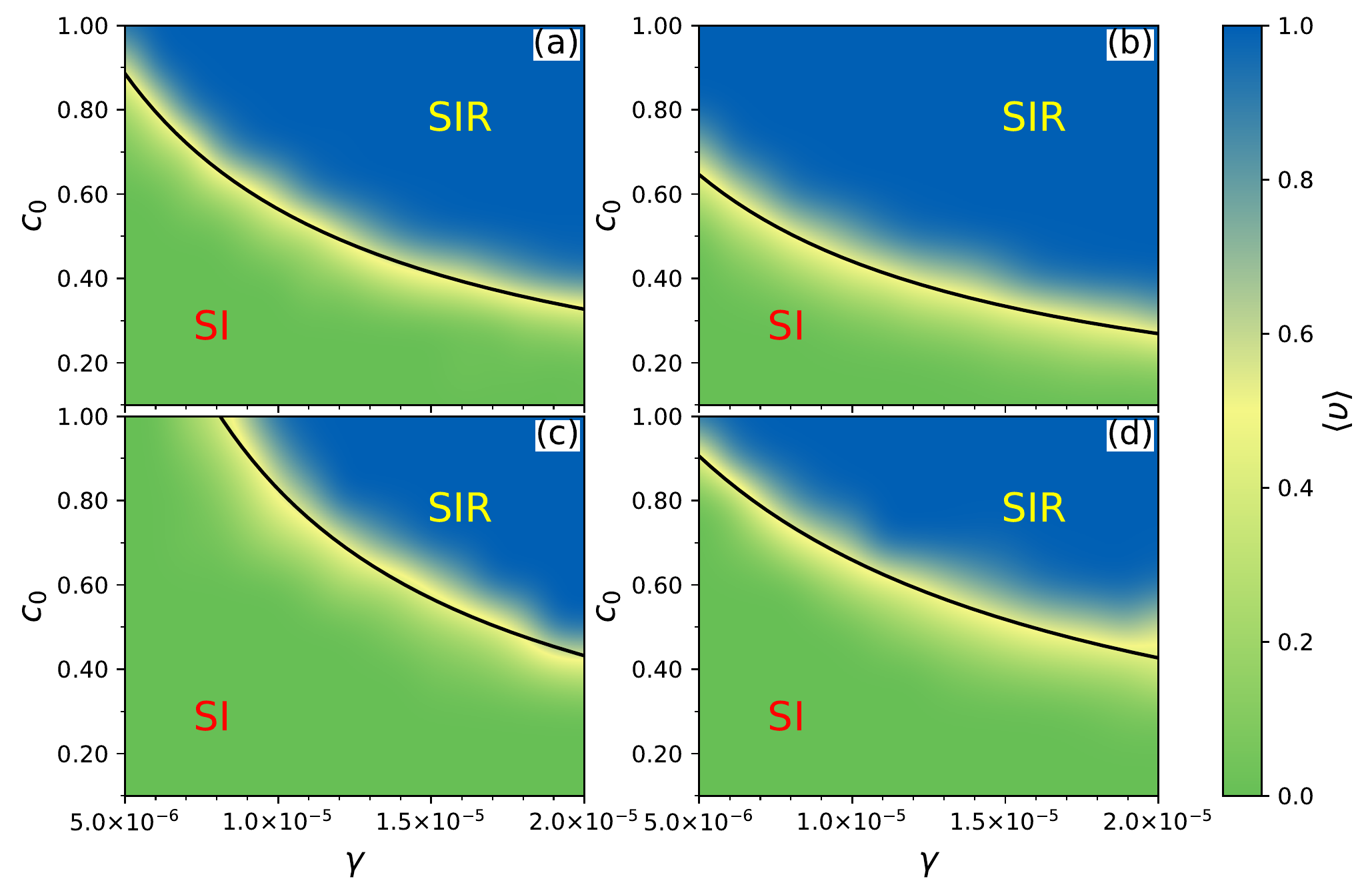}
\caption{
Phase diagrams showing heat maps of $\langle \upsilon\rangle$,
the average fraction of the initial population of $S$ and $C$ that
survive the epidemic,
as a function of initial cleric fraction $c_0$ vs the
probability $\gamma$ of the transition $C+Z\rightarrow C+R$.
Blue indicates SIR behavior in which $Z$ are eliminated, and
green indicates SI behavior in which $S$ and $C$ are eliminated.
In general, as $\gamma$ increases, the SIR behavior emerges at a lower
value of $c_0$.
(a)
Samples of the type shown in Figs.~\ref{fig:1} to \ref{fig:3}
with $\alpha=5\times10^{-6}$ and $\beta=1 \times 10^{-5}$.
(b) Samples with the same $\beta=1 \times 10^{-5}$ where $\alpha$,
the probability for $C+Z\rightarrow Z+Z$, has been halved to
$\alpha=2.5\times10^{-6}$.
(c) Samples with the same $\beta=1 \times 10^{-5}$ in which
$\alpha$ has been doubled to
$\alpha=1\times10^{-5}$.
(d) Samples with
the same $\alpha=5\times10^{-6}$ in which $\beta$, the probability
for $S+Z\rightarrow Z+Z$, is doubled 
to $\beta=2\times10^{-5}$.  
The solid lines in the figures are fits
of the form $c_0 \propto a(\gamma+ b)^{-1}$
where (a) $a=7.77\times 10^{-6}$ and $b=3.781 \times 10^{-6}$,
(b) $a = 6.912\times 10^{-6}$ and $b=5.696\times 10^{-6}$,
(c) $a=9.056\times 10^{-6}$ and $b=9.498\times 10^{-7}$, 
and
(d) $a=1.212\times 10^{-5}$ and $b=8.381 \times 10^{-6}$.
}
\label{fig:4}
\end{figure}

In Fig.~\ref{fig:4} we plot phase diagrams of $\upsilon$ as a function
of $c_0$, the initial cleric fraction,
versus $\gamma$, the probability of the transition
$C+Z\rightarrow C+R$.
Each diagram contains 160 points, and each point is averaged
over 50 different initial realizations.
In the blue region, $\upsilon$ is high and we find SIR-like behavior where
$S$ and $C$ survive while $Z$ are eliminated,
while in the green region, $\upsilon$ is low and the system is SI-like,
with $Z$ persisting to the end of the epidemic and all of the $S$ and $C$
vanishing.
Figure~\ref{fig:4}(a) shows the phase diagram for samples with
$\alpha=5\times10^{-6}$ and $\beta=1 \times 10^{-5}$, as in
Figs.~\ref{fig:2} and \ref{fig:3}.
At higher $\gamma$, the zombies are more effectively healed by the clerics,
and the initial fraction $c_0$ of $C$ needed to produce
SIR-like behavior drops to lower values, as shown by the
solid line which is a fit of the SI-SIR transition to the form
$c_0 \propto a(\gamma+b)^{-1}$. 
For a simple way to understand the general form of this curve, consider
the early time behavior of an individual $Z$ particle.
As it moves, the $Z$ encounters a $C$ with probability $c_0$ and
an $S$ with probability $1-c_0$. The $Z$ always survives an encounter
with $S$, but it only survives an encounter with $C$ with probability
$1-\gamma$. Thus, the probability that the $Z$ survives is
$Z_{\rm survive}=(1-\gamma)c_0+(1-c_0)$ and the probability that the $Z$
is destroyed by turning into an $R$ is
$Z_{\rm destroy}=\gamma c_0$. At the SI-SIR transition, we
have $Z_{\rm survive}=Z_{\rm destroy}$, meaning that the
transition line is expected to fall at $c_0=0.5(\gamma)^{-1}$.

The actual location of the SI-SIR transition line is affected by the
values of $\alpha$ and $\beta$ because these control
the way in which the populations of $S$, $C$, $Z$, and $R$ evolve
over time.
If we cut the probability $\alpha$ of the $C+Z\rightarrow Z+Z$ transition
in half to $\alpha=2.5\times10^{-6}$,
the phase diagram in Fig.~\ref{fig:4}(b) indicates that
the SI-SIR transition line shifts to lower values of $c_0$ since it
becomes more difficult for the $Z$ to eliminate all of the $C$.
If we instead double $\alpha$ to $\alpha=1\times10^{-5}$, as in
Fig.~\ref{fig:4}(c), we reach the limit in which $\alpha=\beta$ and the
$S$ and $C$ particles are both equally likely to be infected upon
encountering a $Z$.
Here, not only does the SI-SIR transition line shift to
higher $c_0$, but for small values of $\gamma$ only SI behavior can occur
even if the entire population apart from the zombie index
cases is initialized to state $C$.
If we leave $\alpha$ unchanged but double $\beta$, the probability
of the $S+Z\rightarrow Z+Z$ transition, to
$\beta=2\times 10^{-5}$, Fig.~\ref{fig:4}(d) shows
that at low $\gamma$, the location of the SI-SIR transition does not change
very much, but at higher $\gamma$, it shifts to higher $c_0$.

\begin{figure*}
\includegraphics[width=0.75\textwidth]{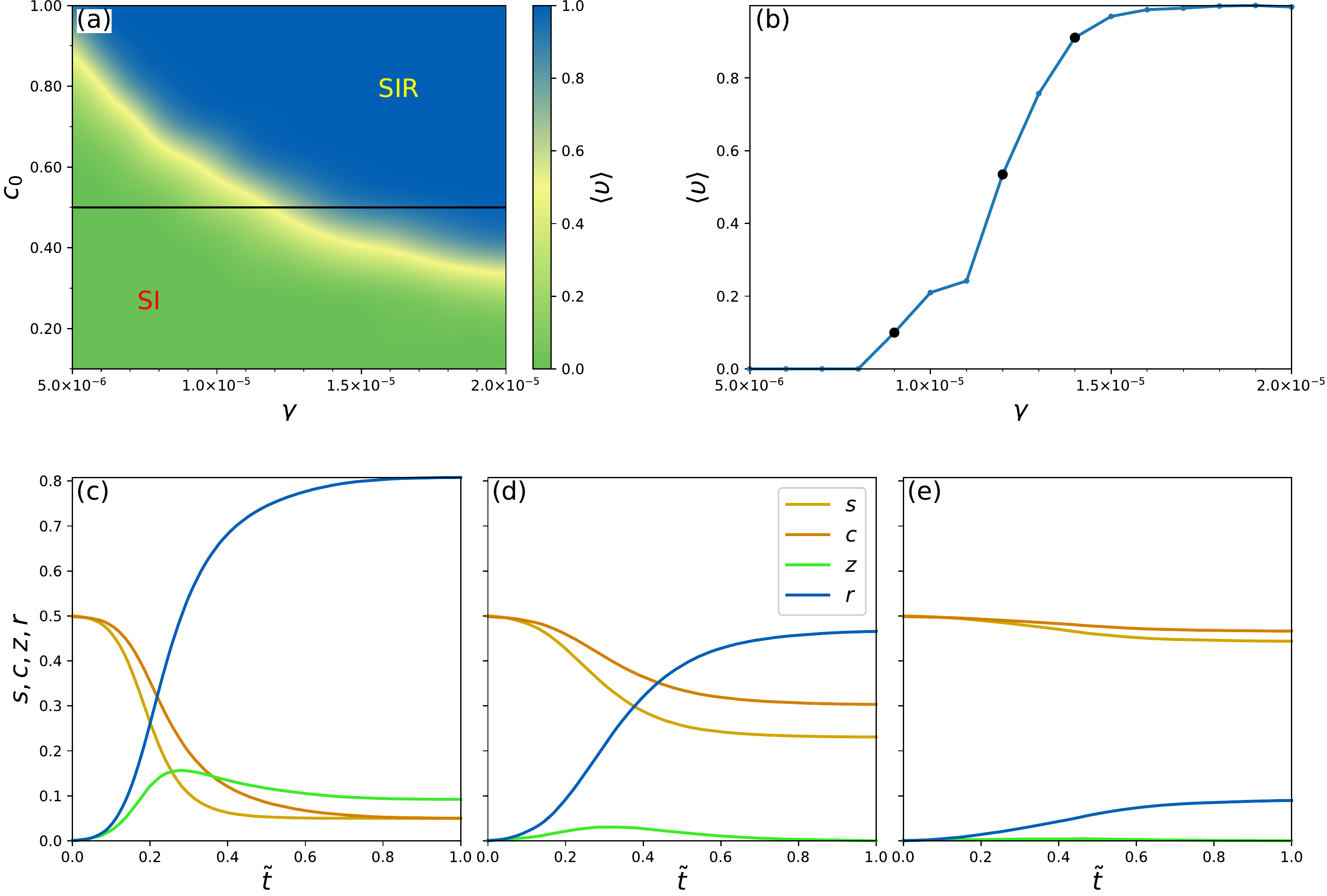}
\caption{
(a) The phase diagram with a heat map of $\langle \upsilon\rangle$ as
a function of $c_0$ vs $\gamma$ from Fig.~\ref{fig:4}(a) with
$\alpha=5 \times 10^{-6}$ and $\beta=1 \times 10^{-5}$. 
(b) A horizontal slice
of $\langle \upsilon\rangle$ vs $\gamma$ taken
at $c_0=0.5$ along the black line in panel (a).
(c,d,e) Epidemic curves averaged over 50 runs
taken at the black points in panel (b) showing
$s$ (yellow), $c$ (orange), $z$ (green), and $r$ (blue)
vs the
rescaled time $\tilde{t}=t/t_d$.
(c) At $\gamma=9\times 10^{-6}$, SI behavior occurs 90\% of the time,
so the averaged values of $s$ and $c$ do not reach zero but are lower
than the averaged value of $z$.
(d) At  $\gamma = 1.2\times 10^{-5}$,
all runs are in the SIR regime and on average 50\% of the
population is never infected.
(e) At $\gamma=1.4 \times 10^{-5}$,
the clerics become more effective at reducing the impact of the
epidemic, and on average 90\% of the population is never infected.
}
\label{fig:5}
\end{figure*}

In order to illustrate some representative averaged epidemic
curves,
in Fig.~\ref{fig:5}(a) we reproduce
the phase diagram of Fig.~\ref{fig:4}(a) for
$\alpha=5\times 10^{-6}$ and $\beta=1 \times 10^{-5}$
with a black line indicating the location of a horizontal cut.
Figure~\ref{fig:5}(b) shows $\langle \upsilon\rangle$ versus
$\gamma$ at the cut location of
$c_0=0.5$.
When $\gamma < 9\times 10^{-6}$, there are no realizations in which
SIR behavior occurs; instead, the $Z$ always wipe out all of the $S$
and $C$. Similarly,
for  $\gamma > 1.1 \times 10^{-5}$, there are no realizations in which SI
behavior occurs, and the $Z$ are always fully eliminated.
The kink in the curve marks the transition to fully SIR behavior.
The value of $\langle \upsilon\rangle$ indicates how effective the
clerics are at suppressing the epidemic.
When $\langle \upsilon\rangle$ increases,
it means that a greater
fraction of the population was never infected by the disease.
For $\gamma$ just above the
transition into fully SIR behavior, over 75\% of the population still
becomes infected before the zombies are eliminated, whereas for
higher $\gamma$, the majority of the population is able to avoid
becoming infected.

For the three points highlighted in black in Fig.~\ref{fig:5}(b),
we show averaged epidemic curves with $s$, $c$, $z$, and $r$ plotted as
a function of normalized time $\tilde{t}=t/t_d$
in Figs.~\ref{fig:5}(c,d,e).
For $\gamma = 9\times 10^{-6}$
in Fig.~\ref{fig:5}(c),
we are still in the SI dominated regime
and the $z$ curve is higher than the $s$ and $c$
curves. Although in any individual run we either have $z=0$ or $s=c=0$,
for the ensemble average $s$ and $c$ are finite
since SIR behavior
emerges 10\% of the time.
Since we are working at $c_0=0.5$, we have $s=c$ at the beginning of
the epidemic, and although $s$ drops more rapidly than $c$ as the
epidemic progresses, by the end of the epidemic
$s \approx c$, due in large part to the many SI runs for which
$s=c=0$.
In Fig.~\ref{fig:5}(d) at $\gamma = 1.2\times 10^{-5}$,
all 50 simulations are in the SIR regime so that $z=0$
at the end of the epidemic, while the final value of
$r \approx 0.5$
shows that on average half of the population becomes infected
before the zombies are extinguished.
Since we have $\beta=2\alpha$, the value of $s$ drops approximately
twice as fast as the value of $c$ at early times in the epidemic,
but as the supply of $Z$ is depleted through healing by the clerics,
both $s$ and $c$ reach a plateau, and in the final state
$c>s$.
For $\gamma=1.9\times 10^{-5}$ in Fig.~\ref{fig:5}(e),
well within the SIR regime,
$z$ remains quite small throughout the epidemic. Although
we still find $c>s$ at the end of the epidemic, both quantities have
dropped only slightly from the original levels and are not very
different from each other, and 90\% of the population is able to
avoid becoming infected.

\begin{figure*}
\includegraphics[width=0.75\textwidth]{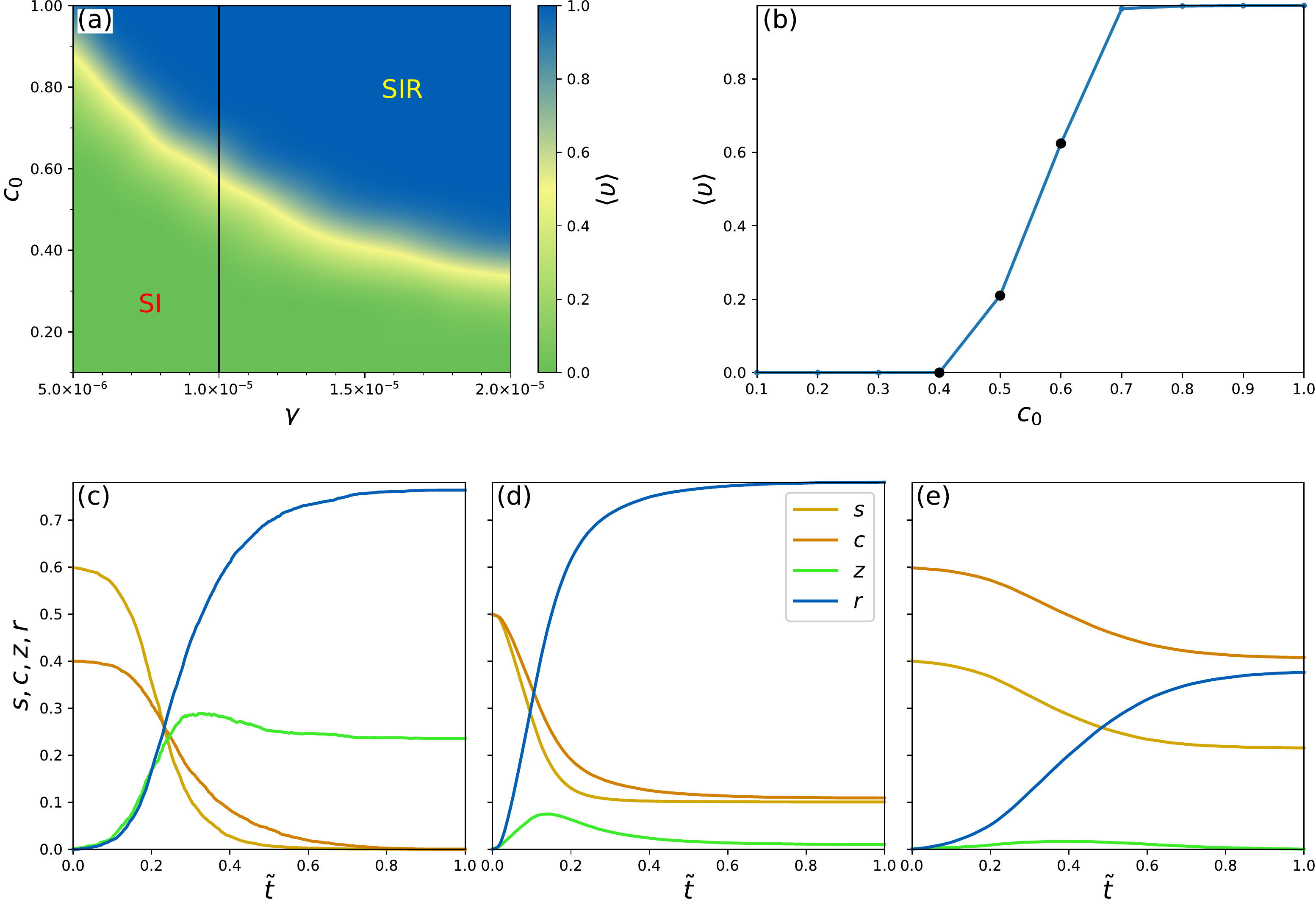}
\caption{The phase diagram with a heat map of $\langle \upsilon\rangle$ as
a function of $c_0$ vs $\gamma$ from Fig.~\ref{fig:4}(a) with
$\alpha=5\times 10^{-6}$ and $\beta=1 \times 10^{-5}$.
(b) A vertical slice of $\langle \upsilon\rangle$ vs $c_0$ taken at
$\gamma=1 \times 10^{-5}$ along the black line in panel (a).
(c,d,e) Epidemic curves averaged over 50 runs
taken at the black points in panel (b) showing $s$ (yellow), $c$ (orange),
$z$ (green), and $r$ (blue) vs $\tilde{t}$.
(c) At $c_0 = 0.4$, only SI behavior occurs.
(d) At $c_0 = 0.5$, we find mixed behavior, with
an SI response occurring 36\% of the time and an SIR
response appearing in the remaining 64\% of runs.
(e) At $c_0= 0.6$, all runs are in the SIR regime.
}
\label{fig:6}
\end{figure*}

As shown in Fig.~\ref{fig:6}(a), 
we next consider a vertical cut at $\gamma=1\times 10^{-5}$ from the
phase diagram in Fig.~\ref{fig:4}(a) for $\alpha=5 \times 10^{-6}$
and $\beta=1 \times 10^{-5}$.
In Fig.~\ref{fig:6}(b) we plot $\langle \upsilon \rangle$ versus $c_0$
along this cut.
For $c_0<0.5$, all of the realizations are in the SI regime and
the $Z$ prevail, while
for $c_0\geq 0.6$, all of the realizations are in the SIR regime and
there are no $Z$ remaining at the end of the epidemic.
The black points in Fig.~\ref{fig:6}(b)
correspond to the values of $c_0$ at which the averaged
epidemic curves in Figs.~\ref{fig:6}(c,d,e) were obtained.
At $c_0=0.4$ in the SI regime,
Fig.~\ref{fig:6}(c) shows that at the end of the epidemic,
$s=c=0$ and
the average fraction of zombies is $z=0.28$.
When $c_0=0.5$ in Fig.~\ref{fig:6}(d),
the system is
in the SI regime 36\% of the time, so that the final
value of $z$ is greater than zero.
Although $c$ and $s$
approach each other toward the end of the epidemic,
we find that $c>s$ by a small amount since the $s=c=0$ behavior
of the SI regime is no longer dominant.
In Fig.~\ref{fig:6}(e), for $c_0=0.6$ the system is
fully in the SIR regime,
and throughout the epidemic we find not only that $c>s$ but that the
difference between $c$ and $s$ remains constant.
This is an indication of the importance of the stochastic diffusive
process that occurs in our model in order to permit $Z$ to come into
contact with $S$ or $C$. For $c_0=0.4$ in Fig.~\ref{fig:6}(c), at early
times in the epidemic a $Z$ encounters an $S$ 60\% of the time but a
$C$ only 40\% of the time. Since $S$ are twice as likely as $C$ to be
infected, $s$ drops much more rapidly than $c$ in this regime.
When $c_0$ is increased to $c_0=0.5$ in Fig.~\ref{fig:6}(d),
a $Z$ is equally likely to encounter an $S$ or a $C$ at early times, and
we see that the doubled infection probability causes
$s$ to drop about twice as fast as $c$, as also shown in
Fig.~\ref{fig:5}(c,d,e). Further increasing $c_0$ to $c_0=0.6$ in
Fig.~\ref{fig:6}(e) means that at early times a $Z$ encounters a $C$
60\% of the time and an $S$ only 40\% of the time. Since the $C$ are
more resistant to infection, the relative fraction of $C$ and $S$ in
the population remains nearly constant. Increasing $c_0$ even further
produces many short-lived epidemics in which $s$ and $c$ do not change
very much from their initial values.

We can analytically evaluate $\upsilon$ for well mixed systems 
whose dynamics is described through Equations (\ref{eq:SCZR1})-(\ref{eq:SCZR4}).
Using a standard argument (see \cite{miller2012}) and some algebra,
we can show that 
\[
%\left ( \frac{S(t)}{S_0} \right )^\alpha = \left ( \frac{C(t)}{C_0} \right )^\beta
\left ( \frac{s}{s_0} \right )^\alpha = \left ( \frac{c}{c_0} \right )^\beta
\]
This provides us with the opportunity to compute a target for
$\upsilon$:
\[
\upsilon = (s_f + c_0(s_f/s_0)^{\alpha/\beta})/(s_0+c_0) \ .
\]
Failure
to hit that target in simulations is an indication that the homogeneous 
mixing assumption failed.
From the data in Figs.~\ref{fig:5}(c,d,e) and \ref{fig:6}(c,d,e), we
find that the predicted value of $\upsilon$ is higher than the actual
value of $\upsilon$, but that the agreement between predicted and actual
improves as we move deeper into the SIR regime. This could be an
indication that the SIR regime is better mixed than the SI regime,
possibly due to the faster dynamics that tend to occur for SI behavior.

\begin{figure}
\includegraphics[width=0.5\textwidth]{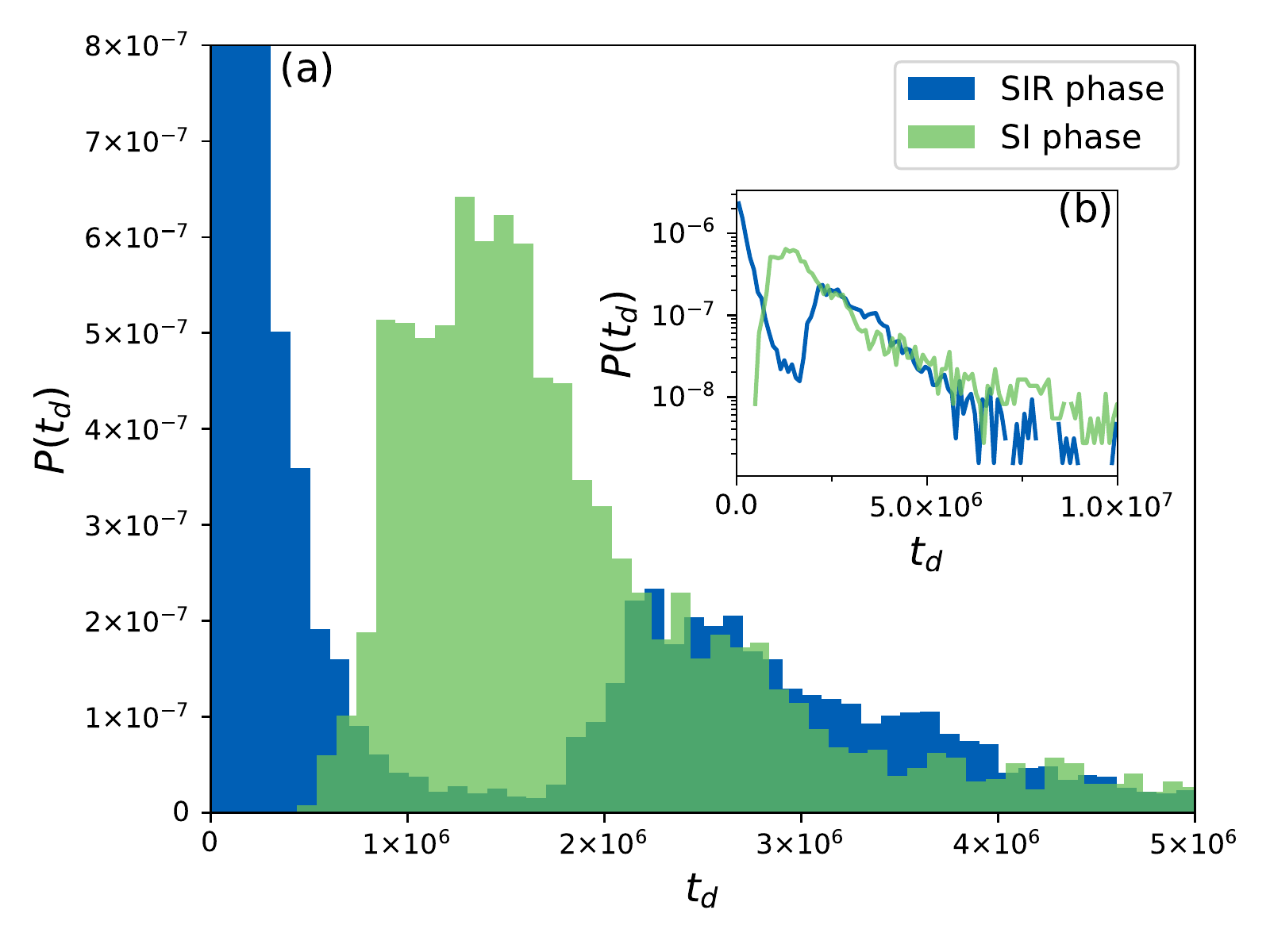}
\caption{The distribution $P(t_d)$ of epidemic durations for the
runs presented in the phase diagrams of
Fig.~\ref{fig:4}.
Blue: runs in which the final state was in the SIR regime
with all of the $Z$ eliminated. Green: runs in
which the final state was in the SI regime with a finite population
of $Z$ surviving at the end of the epidemic.
Inset: The same data plotted on a log-linear scale.
}       
\label{fig:7}
\end{figure}

In Fig.~\ref{fig:7} we plot the distribution $P(t_d)$ of the 
duration $t_d$ of the individual epidemics for
the runs in all of the phase diagrams
in Fig.~\ref{fig:4}.
The data is split into two distributions, with the first for simulations that
ended in the SI regime with a finite number of $Z$ remaining,
and the second for simulations that ended in the SIR regime with no $Z$
remaining.
For the SI case, there are no epidemics of short duration. This is
because all $C$ and $S$ must be eliminated in the SI regime, and the
elimination process requires a minimum amount of time to occur.
In the inset we show the same data on a log-linear scale, indicating that
some of the SI epidemics last for an extremely long time before reaching
a final state.
These lengthy epidemics occur for values of $c_0$ and $\gamma$ at which
the behavior is evenly split between SI and SIR on average.
There is also a peak in $P(t_d)$ near $t_d=1.5 \times 10^{5}$ simulation
time steps.
In the SIR regime, there is a large peak in $P(t_d)$ at small $t_d$
corresponding to failed outbreaks in which the $C$ can rapidly
encounter and cure the small number of $Z$ present at early times
before the epidemic gets going. This is followed by a gap similar to what
we observed previously in SIR simulations \cite{Forgacs22}, and
then by a second peak representing epidemics that involve a substantial
portion of the population. Here we find that if the epidemic
in the SIR regime is able
to become established, it lasts longer than the typical epidemic in
the SI regime, but that there is a high probability for the SIR epidemic
to be extinguished before it can become established.

\section{Discussion}
As we noted earlier, although we have cast our SCZR model
in terms of zombies and clerics,
it could also be rephrased so that the zombies are
disease-spreading individuals
that cannot spontaneously recover from the disease they have
caught, and the clerics are medical care providers who can
cure the infected individuals or at least render them
non-infectious.
In this picture, when we take $\alpha<\beta$ but $\alpha>0$, this
would mean that the medical care providers are more careful than the
general population and take more precautions against becoming
infected, but that they are not immune from becoming
infected.
The transition between SI and SIR behavior is significant because
it indicates that by introducing a larger number of medical care
providers (increasing $c_0$) or giving the medical care providers
more effective treatment protocols (increasing $\gamma$), the
disease can be prevented from entering the SI regime in which the
entire population winds up getting infected eventually, and can instead
be held in the SIR regime, ideally in the limit where $t_d$ is short and
the epidemic never becomes established in the population.
Some of the next steps for our SCZR  model would be to consider the effect
of adding fixed spatial heterogeneity such as quenched disorder.
For example, the $C$ might be confined to only certain regions of
the system, as in real world scenarios where impassable terrain or
military blockades are present. Other situations include considering
the case where the $R$ are not epidemiologically inert but can produce
infection at greatly reduced rates $\beta^{\prime} \ll \beta$ and
$\alpha^{\prime}\ll\alpha$, to represent situations in which the medical
care givers only reduce the infectiousness rather than fully eliminating
it.
Active matter models in general also readily allow other effects to
be captured,
such as introducing a small fraction of very active particles with
increased motor force $F_M$ embedded in a population of reduced
mobility or much smaller $F_M$ in order to
represent different types of mobility patterns in 
social systems.

Another question that could be explored with the SCZR model
is what is the nature of the transition from the SI to the SIR
regime. Although the transition is somewhat sharp in our phase
diagrams, it may be only a crossover.
Note that in the limit $c_0=1$, the SCZR model
becomes equivalent to the SZR model
of Ref.~\cite{Alemi15}. In this limit, Fig.~\ref{fig:4} shows that
for certain parameter regimes there
is still a transition from SI to SIR behavior; however, it is much
more intuitive from a medical intervention point of view
to tune between the two regimes using the
$c_0$ and $\gamma$ parameters of the SCZR model than by using 
the parameter $\alpha$ (which is written as $\beta$ in the SZR model).
Epidemic models show various
types of critical phenomena associated with
directed percolation transitions \cite{Grassberger83,Tome10};
however, such transitions can be 
screened or modified
by the introduction of quenched disorder \cite{Mukhamadiarov22},
so we expect that there could be
various types of critical behavior in our system. 

\section{Summary}

We have introduced a model for epidemics that we call the
Susceptible-Cleric-Zombie-Removed or SCZR model,
and we demonstrate the use of this model
with active matter run-and-tumble particles.
In the SCZR model, the infectious agents are the zombies, and there is no
spontaneous recovery. There is an initial population of susceptibles and
clerics. With probability $\alpha$ for clerics and $\beta$ for
susceptibles, interaction with a zombie causes infection into the
zombie state,
while with probability $\gamma$, a cleric
interacting with a zombie causes the zombie to enter an
epidemiologically inert recovered state.
We show that by varying the initial density of clerics or their healing
rate $\gamma$, we can tune the SCZR model between SI and SIR regimes.
If the initial cleric density or the healing rate $\gamma$ is low,
the zombies eliminate all of the clerics and susceptibles to give SI
behavior, while if the initial cleric density or healing rate $\gamma$ is
high enough, the clerics are able to heal all of the zombies and
SIR behavior emerges.
Our model has implications for real world diseases where
infections are lifelong and spontaneous recovery does not occur, but
where medical intervention can produce recovery or at least drive the
rate of infectiousness to zero.
One example of this type of disease is the human immunodeficiency virus (HIV).
In this case, the zombies would be infected persons and the clerics would
represent medical caregivers that can provide treatment.
The SCZR model could provide a good staring point for
creating new types of epidemic models where treatment is needed
for recovery and there are finite or limited treatment resources available. 

\smallskip

\begin{acknowledgments}
This work was supported by the US Department of Energy through
the Los Alamos National Laboratory.  Los Alamos National Laboratory is
operated by Triad National Security, LLC, for the National Nuclear Security
Administration of the U. S. Department of Energy (Contract No. 892333218NCA000001).
NH benefited from resources provided by the Center for Nonlinear Studies (CNLS).
PF and AL were supported by a grant of the Romanian Ministry of Education
and Research, CNCS - UEFISCDI, project number
PN-III-P4-ID-PCE-2020-1301, within PNCDI III.
\end{acknowledgments}

\bibliography{mybib}

\vfill\eject
\setcounter{figure}{0}

\end{document}